\documentclass[11pt,a4paper,conference]{IEEEtran}

\usepackage{graphicx}
\usepackage{amssymb}
\usepackage{caption}
\usepackage{subcaption}
\usepackage{url}
\usepackage{amsmath}
\usepackage{float}
\usepackage{caption}
\usepackage{color}

\def \F{\vec{F}}

\def\U {{{\bf u}}}
\def\F {{{\bf F}}}

\def\Y {{{\bf Y}}}

\def\div {{\nabla \cdot}}

\graphicspath{{./Images/}}

\begin{document}

%
%

\title{Pulsing corals: A story of scale and mixing}

%
%

\author{\IEEEauthorblockN{Julia E. Samson\IEEEauthorrefmark{2},
Nicholas A. Battista\IEEEauthorrefmark{3},
Shilpa Khatri\IEEEauthorrefmark{4}, and 
Laura A. Miller\IEEEauthorrefmark{1}}
\IEEEauthorblockA{\IEEEauthorrefmark{2}Dept. of Biology\\
University of North Carolina at Chapel Hill, Chapel Hill, NC, USA\\
jesamson@live.unc.edu}
\IEEEauthorblockA{\IEEEauthorrefmark{3}Dept. of Mathematics and Statistics\\
The College of New Jersey, Ewing, NJ, USA\\
battistn@tcnj.edu}
\IEEEauthorblockA{\IEEEauthorrefmark{4}Applied Mathematics Unit, School of Natural Sciences\\
University of California Merced, Merced, CA, USA\\
skhatri3@ucmerced.edu}
\IEEEauthorblockA{\IEEEauthorrefmark{1} Dept. of Biology, Dept. of Mathematics\\
University of North Carolina at Chapel Hill, Chapel Hill, NC, USA\\
lam9@unc.edu}
}

\maketitle

%
%

\begin{abstract}
Effective methods of fluid transport vary across scale. A commonly used dimensionless number for quantifying the effective scale of fluid transport is the Reynolds number, $Re$, which gives the ratio of inertial to viscous forces. What may work well for one $Re$ regime may not produce significant flows for another. These differences in scale have implications for many organisms, ranging from the mechanics of how organisms move through their fluid environment to how hearts pump at various stages in development. Some organisms, such as soft pulsing corals, actively contract their tentacles to generate mixing currents that enhance photosynthesis. Their unique morphology and intermediate scale where both viscous and inertial forces are significant make them a unique model organism for understanding fluid mixing. In this paper, $3D$ fluid-structure interaction simulations of a pulsing soft coral are used to quantify fluid transport and fluid mixing across a wide range of $Re$. The results show that net transport is negligible for $Re<10$, and continuous upward flow is produced for $Re \geq 10$.
\end{abstract}


%
%

\section{Introduction}

Efficient fluid transport is not only dependent on the method of movement, but also the fluid's physical properties and scale. While one mechanism for transport may work efficiently at one scale, it may not work at all at another. For example, reciprocal motion of a fish's caudal fin may not produce adequate forward propulsion if the fish is put into a considerably more viscous fluid than water. If the viscosity is high enough, the fish might not swim at all, and every fin stroke will yield no net transport. The fact that reciprocal motions do not generate net movement at small scales is famously known as the Scallop Theorem \cite{Purcell:1977}. The Reynolds number, $Re$ is a dimensionless quantity of the ratio of inertial to viscous forces in a fluid used to compare fluid transport across scales. For a fluid of density, $\rho$, dynamic viscosity, $\mu$, and some characteristic length and frequency scale, $L$ and $f$, respectively, a frequency based $Re$ may be defined as
\begin{equation}
    \label{Re} Re = \frac{\rho L^2 f}{\mu}.
\end{equation}
For sufficiently low $Re$ for a Newtonian fluid in a large domain, it is necessary to use a non-reciprocal motion to transport fluid. One common example is the use of a rotating flagella as in the case of many bacteria and sperm \cite{Dillon:2006,Olson:2011}. Beyond locomotion of an organism, there are many other applications of fluid transport within biological systems. Examples include the generation of feeding currents \cite{Hamlet:2012}, the generation of flow for oxygen and nutrient transport \cite{Truskey:2004}, the internal pumping of fluids as in the case of the cardiovascular system \cite{Peskin:1996}, flows generated for filtering \cite{Cheer:1987}, and flows for photosynthetic enhancement \cite{Shapiro:2014}. As in the case for locomotion, different mechanisms of pumping and feeding may only be effective over some range of $Re$ \cite{Baird:Re,Holzman:Re}.

\begin{figure}
\centering
\includegraphics[width=0.48\textwidth]{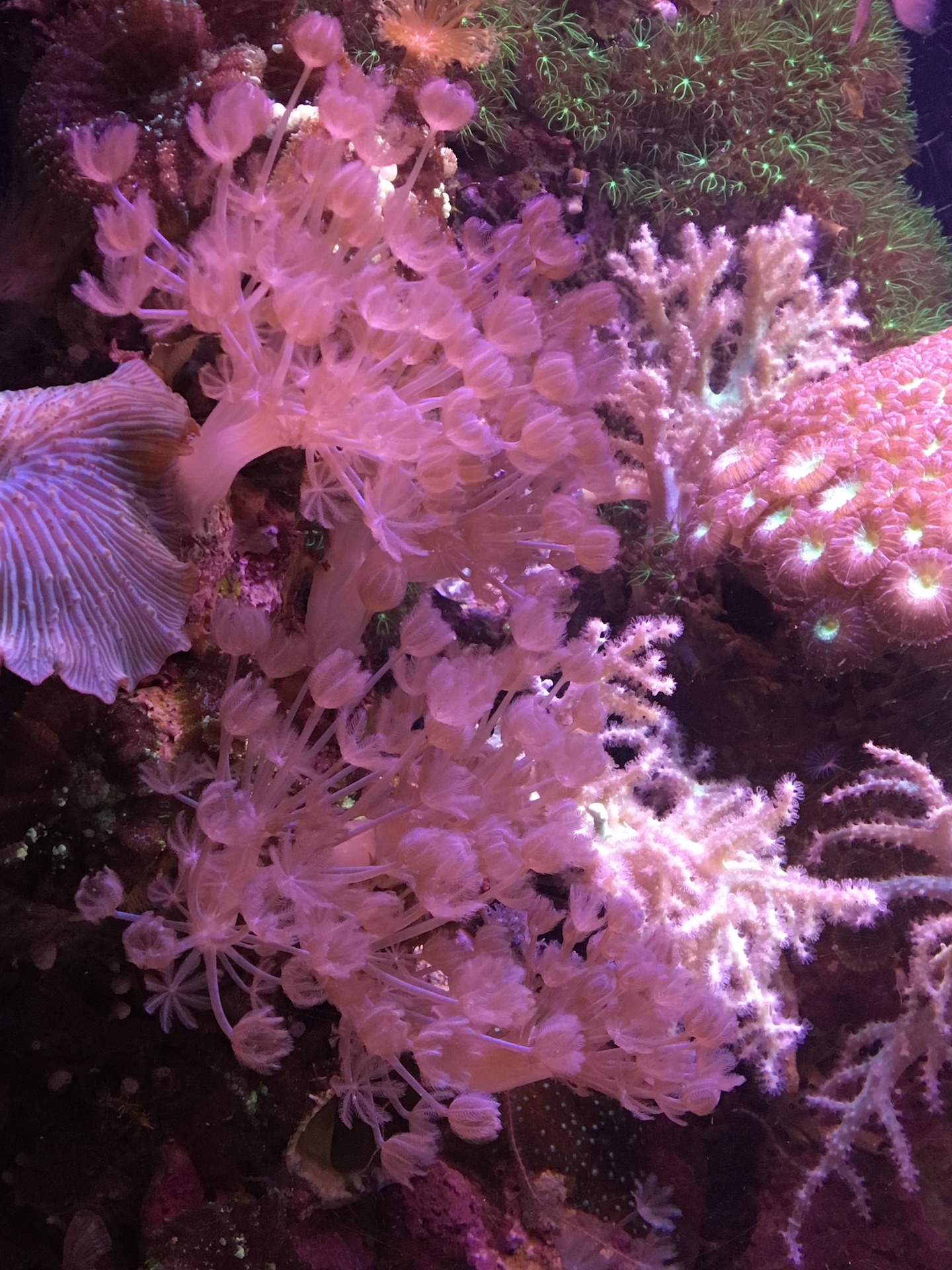}
\caption{Xenia polyps from the Underwater Observatory, Eilat, Israel.}
\label{fig:Xenia_Colony}
\end{figure}

One example of an organism that moves for the purpose of enhancing exchange is the pulsing soft coral, such as \textit{Heteroxenia fuscescens}. These soft bodied corals live in colonies and actively contract their tentacles. Each individual polyp contains eight feather-like, pinnate tentacles, see Figure \ref{fig:Xenia_Colony}. Each stalk is approximately $5\ cm$ long, and the colony can grow up to $60\ cm$ across \cite{Lieske:2004}. 

The colony does not normally pulse in sync, but out of phase \cite{Kremien:2013}. On average, each polyp tends to pulse $1$-$2$ times per second. It was not clear how \textit{Xenia} benefit from pulsing until a recent study by Kremien et al. \cite{Kremien:2013} that showed that pulsing enhances photosynthetic rates by up to an order of magnitude. This is done by controlling chemical transport around the colony in comparison to their non-pulsing counterparts.

Most of the fluid dynamic studies of rhythmic pulsing in Cnidarians, such as jellyfish and corals, have focused on locomotion. One interesting exception is the upside down jellyfish (\textit{Cassiopeia}) that pulse while sitting on the sea floor. While they can swim, they typically rest upside down on the substrate and pulse for feeding and nutrient exchange. Since they host zooxanthellae in their tissues, they also photosynthesize \cite{Kaplan:1988,Fitt:1998}. Unlike soft corals that generate exchange currents with their tentacles, upside down jellyfish create flow by actively contracting and relaxing their gelatinous bell. The biologically relevant $Re$ for an upside down jellyfish ranges from $Re=45$ (juvenile) to approximately $Re=450$ (adult) \cite{Hamlet:Thesis} such that they operate in the inertial range ($Re>>1)$, where reciprocal motions are effective. Several experimental and computational investigations have described the fluid dynamics of upside down jellyfish \cite{Hamlet:2012,Hamlet:2012b,Hamlet:2014}.  

Pulsing soft corals also operate in the inertial regime, where $Re\sim 10-100$, see Section \ref{sec:methods}. Although both upside down jellyfish and single coral polyps symmetrically contract their bell or tentacles, respectively, it is anticipated that the differences in scale and morphology will result in different fluid dynamics between the two species. In this paper we investigate the fluid dynamics of one pulsing soft coral over a range of $Re$, using a fully coupled fluid-structure interaction model. We explore the net flow in both vertical and horizontal directions and describe the dynamics of the vortex rings and jets generated across $Re$. 


%
%

\section{Methods}
\label{sec:methods}

The immersed boundary method (IB) \cite{Peskin:2002} was used to solve the fully coupled fluid-structure interaction problem of a pulsing soft coral in an incompressible, viscous fluid. The IB has been successfully applied to a variety of applications in biological fluid dynamics within an intermediate $Re$ regime, e.g., $0.01<Re<1000$, including heart development \cite{Battista:2016a,Battista:2016b}, insect flight \cite{SJones:2015}, swimming \cite{Hoover:2015,Hoover:2017}, and dating and relationships \cite{BattistaIB2d:2016}. A fully parallelized implementation of the IB with adaptive mesh refinement, IBAMR, was used for the simulations described here \cite{BGriffithIBAMR}. More details on IB and IBAMR are found in Appendix \ref{IB_Appendix}.

\begin{figure}
\centering
\includegraphics[width=0.45\textwidth]{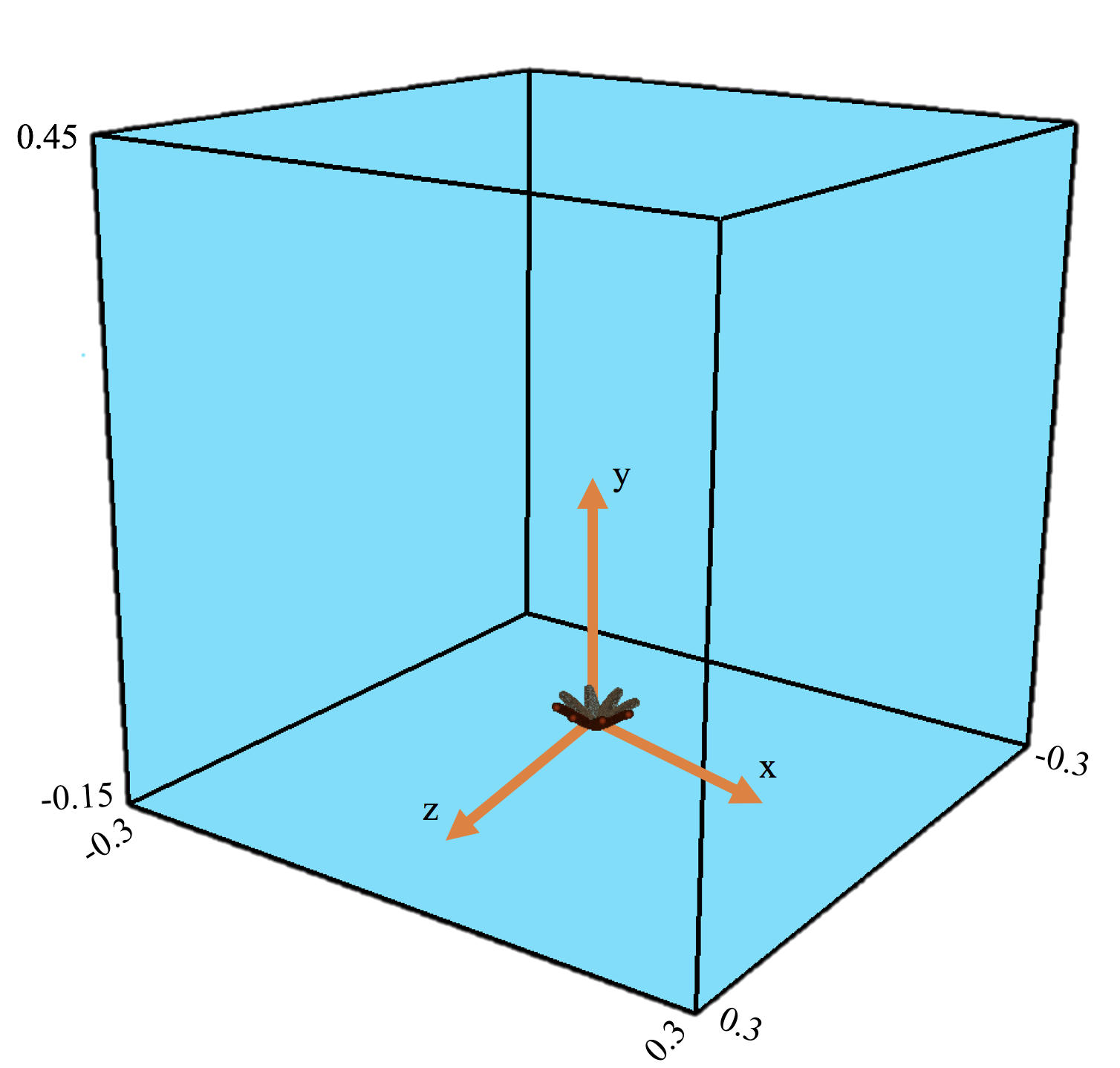}
\caption{The computational domain for a single coral polyp. Note that the boundaries in the $x$- and $z$-directions are periodic. The boundary conditions in the $y$-direction are no slip ($\textbf{u}=0$ at $y=-0.15$ and $y=0.45$) .}
\label{fig:comp_domain}
\end{figure}

All parameter values used in the computational model are given in Table \ref{table:num_param}. A depiction of the computational domain is given in Figure \ref{fig:comp_domain}. Note that periodic boundaries are used in the $x$ and $z$ directions, and no slip conditions are used in the $y$-direction corresponding to a solid boundary on the top and bottom of the domain ($\textbf{u}=0$ at $y=-0.15$ and $y=0.45$). The initial conditions of the fluid are set to zero and there is no ambient flow considered. For a study including ambient flow see the follow up paper in this issue \cite{Battista:coral-flow}. 

The $Re$ is computed using the characteristic length set to the tentacle length and the characteristic frequency set to the pulsation frequency. The fluid density and dynamic viscosity are set to that of sea water. A biologically relevant $Re$ for a pulsing coral with measurements of $f_{coral} = 1/1.9 s^{-1}$, $L_{coral} = 0.0045 m$, $\rho=1023\ kg/m^3$ and $\mu= 0.00096 kg/(ms),$ is 
\begin{equation}
    \label{Re_Xenia} Re = \frac{ \rho f_{coral} L^2_{coral} }{\mu} = 10.66.
\end{equation}
The dynamic viscosity will be varied in the simulations to study a range of $Re$. The range of $Re$ studied here is 0.5, 1, 5, 10, 20, 40, and 80. 

\begin{table}
\begin{center}
\begin{tabular}{| c | c | c | c |}
    \hline
    Parameter               & Variable    & Units        & Value \\ \hline
    Domain Size            & $D$          & m               &  $0.06$               \\ \hline
    Spatial Grid Size      & $dx$         & m               &  $D/1024$               \\ \hline
    Lagrangian Grid Size    & $ds$        & m               &  $D/2048$               \\ \hline
    Time Step Size          & $dt$        & s               &  $1.22\times 10^{-4}$   \\ \hline
    Total Simulation Time    & $T$        & \textit{pulses} &  $10$               \\ \hline
    Fluid Density            & $\rho$     & $kg/m^3$        &  $1000$               \\ \hline
    Fluid Dynamic Viscosity & $\mu$      & $kg/(ms)$       &  \textit{varied}      \\ \hline
    Tentacle Length          & $L_T$      & m               &  $0.0045$               \\ \hline
    Pulsing Period           & $P$        & s               &  $1.9$               \\ \hline
    Target Point Stiffness   & $k_{target}$&$kg\cdot m/s^2$ &  $9.0\times10^{-9}$  \\ \hline
    \hline
    \end{tabular}
    \caption{Numerical parameters used in the three-dimensional simulations.}
    \label{table:num_param}
    \end{center}
\end{table}

\begin{figure}
\centering
\includegraphics[width=0.45\textwidth]{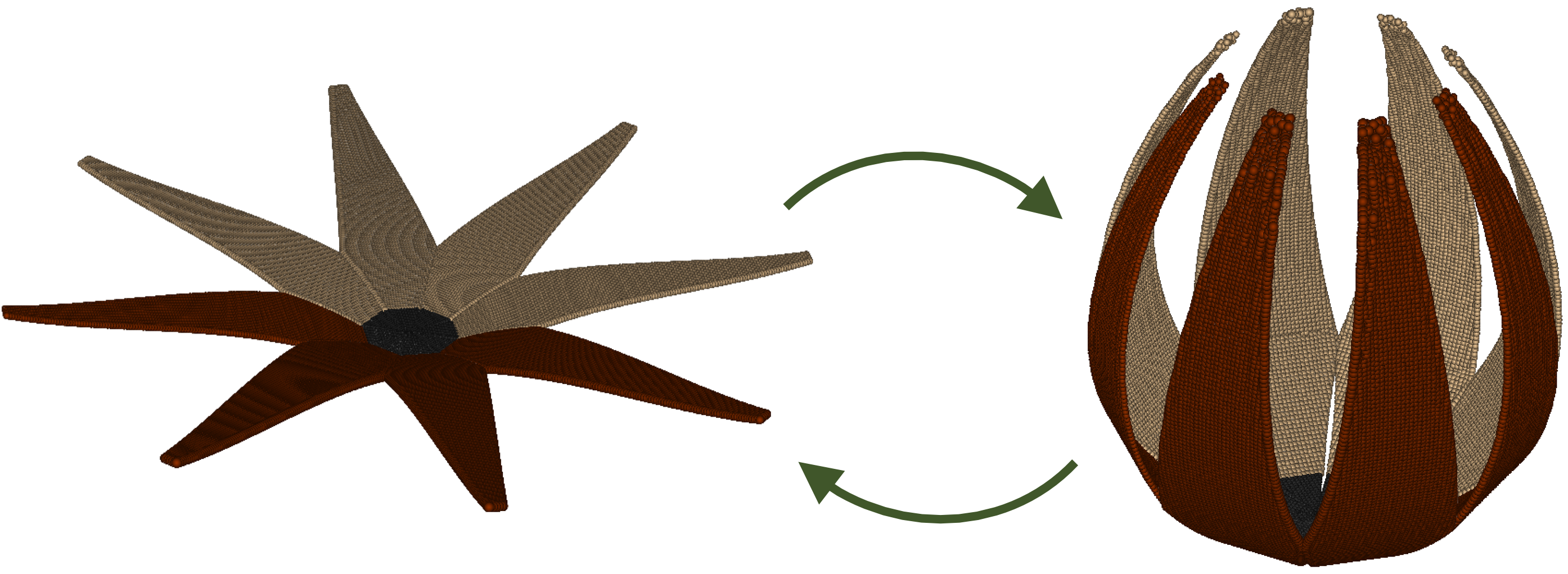}
\caption{A single polyp's pulsation cycle. The coral moves from its relaxed state to an actively contracted state and then relaxes back to its original, open, resting state. The tentacle colors were chosen make each tentacle distinct.}
\label{fig:pulsation_cycle}
\end{figure}

The pulsing motion of the coral was based on kinematics of the live organism. This motion was included in the IB method by tethering the Lagrangian geometry of the tentacles to target points. The target points were moved in a prescribed fashion to best mimic the kinematics. The kinematics were captured by tracking positions along a single tentacle from 5 different coral polyps. These positions were then fit with polynomials and then averaged over the polyps and averaged over multiple pulses. This average behavior of a tentacle is what was used to enforce the prescribed motion of the immersed boundary. For more details on this analysis, see \cite{Samson:coral}. A coral pulsation cycle was divided into the 3 phases, see Figure \ref{fig:pulsation_cycle} and below, 
\begin{enumerate}
    \item [1.] The coral begins with its tentacles all in their open, relaxed state, and then move to a closed, actively contracted state. This takes about 28\% of the pulse cycle.
    \item [2.] From the contracted state, the tentacles relax back to their original open, resting state. The expansion phase takes about 43\% of the pulse cycle.
    \item [3.] The tentacles remain open and at rest for about 29\% of the pulse cycle. 
\end{enumerate}
The process then repeats itself.

%
%
\section{Results}

\begin{figure}
\centering
\includegraphics[width=0.45\textwidth]{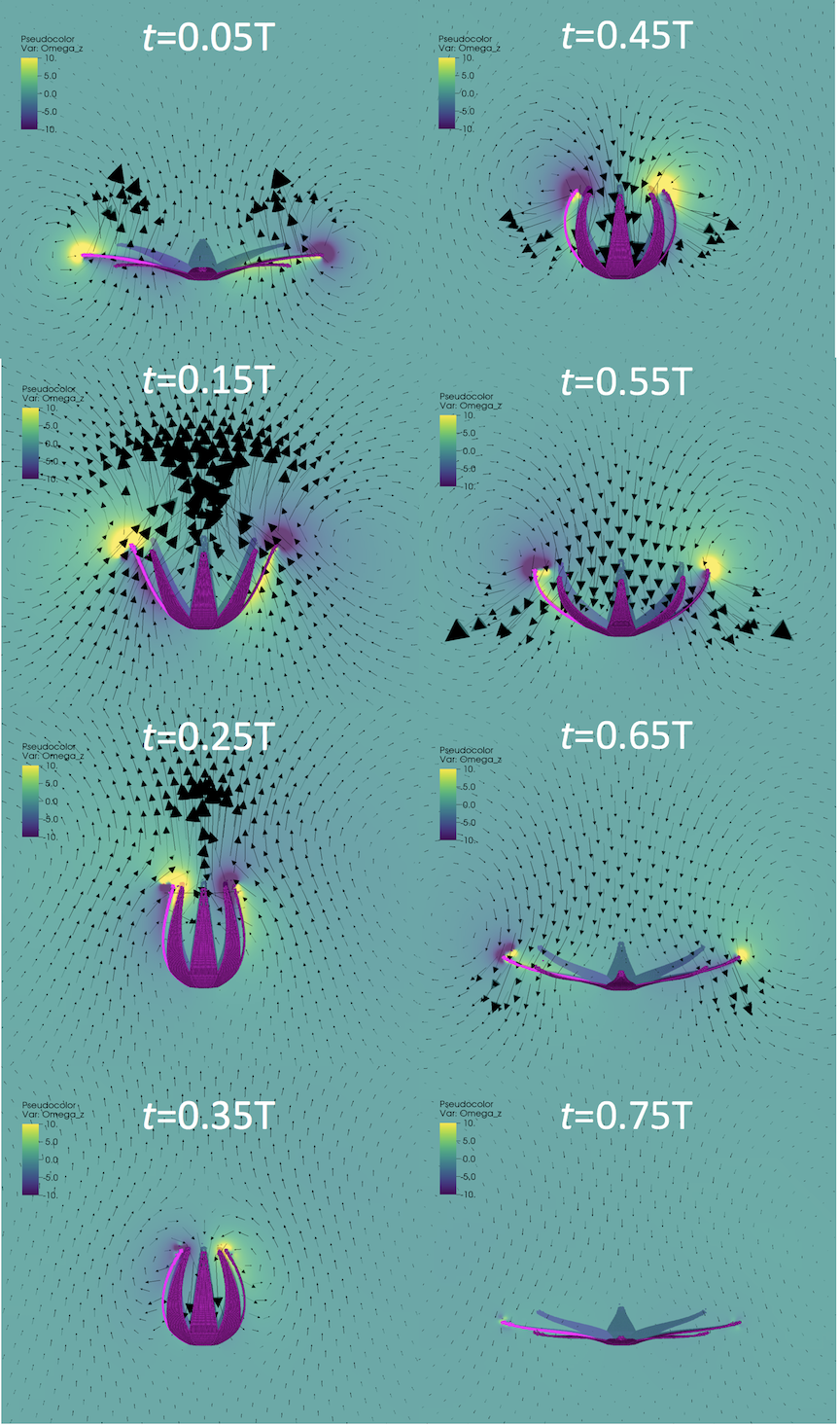}
\caption{The $z$-component of vorticity and the velocity vector field taken on a 2D plane through the central axis of the coral at $Re=0.5$. This $Re$ corresponds to a smaller scale than would be observed in nature. The colormap shows the value of $\omega_z$, the arrows point in the direction of flow, and the length of the vectors correspond to the magnitude of the flow. Shapshots are taken during the fourth pulse at times that are 5\%, 15\%, 25\%, 35\%, 45\%, 55\%, 65\%, and 75\% through the cycle. }
\label{fig:Re_0p5_snapshots}
\end{figure}

\begin{figure}
\centering
\includegraphics[width=0.45\textwidth]{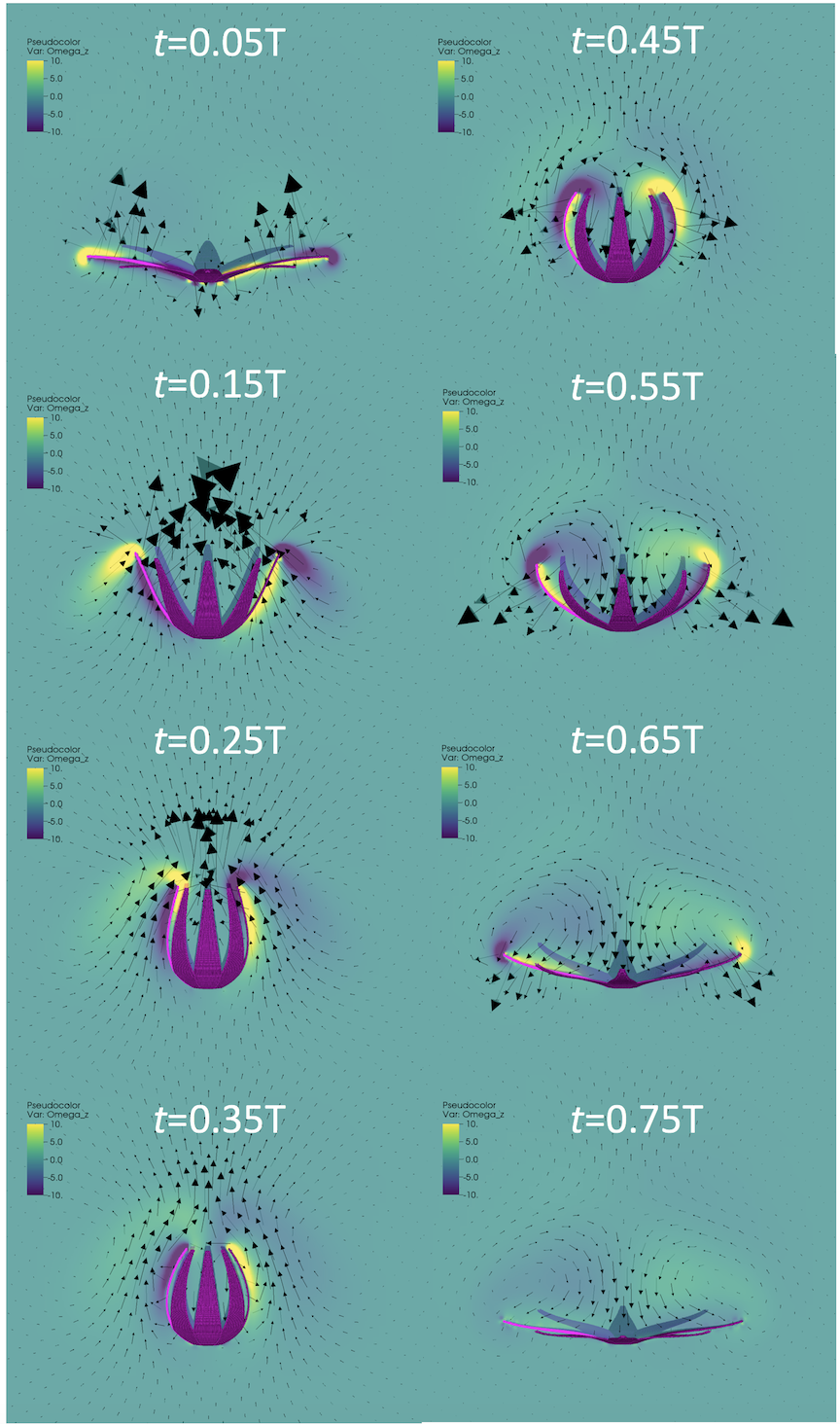}
\caption{The $z$-component of vorticity and the velocity vector field taken on a 2D plane through the central axis of the coral at $Re=10$. This $Re$ corresponds to a typical coral polyp.The colormap shows the value of $\omega_z$, the arrows point in the direction of flow, and the length of the vectors correspond to the magnitude of the flow. Snapshots are taken during the fourth pulse at times that are 5\%, 15\%, 25\%, 35\%, 45\%, 55\%, 65\%, and 75\% through the cycle.}
\label{fig:Re_10_snapshots}
\end{figure}

\begin{figure}
\centering
\includegraphics[width=0.45\textwidth]{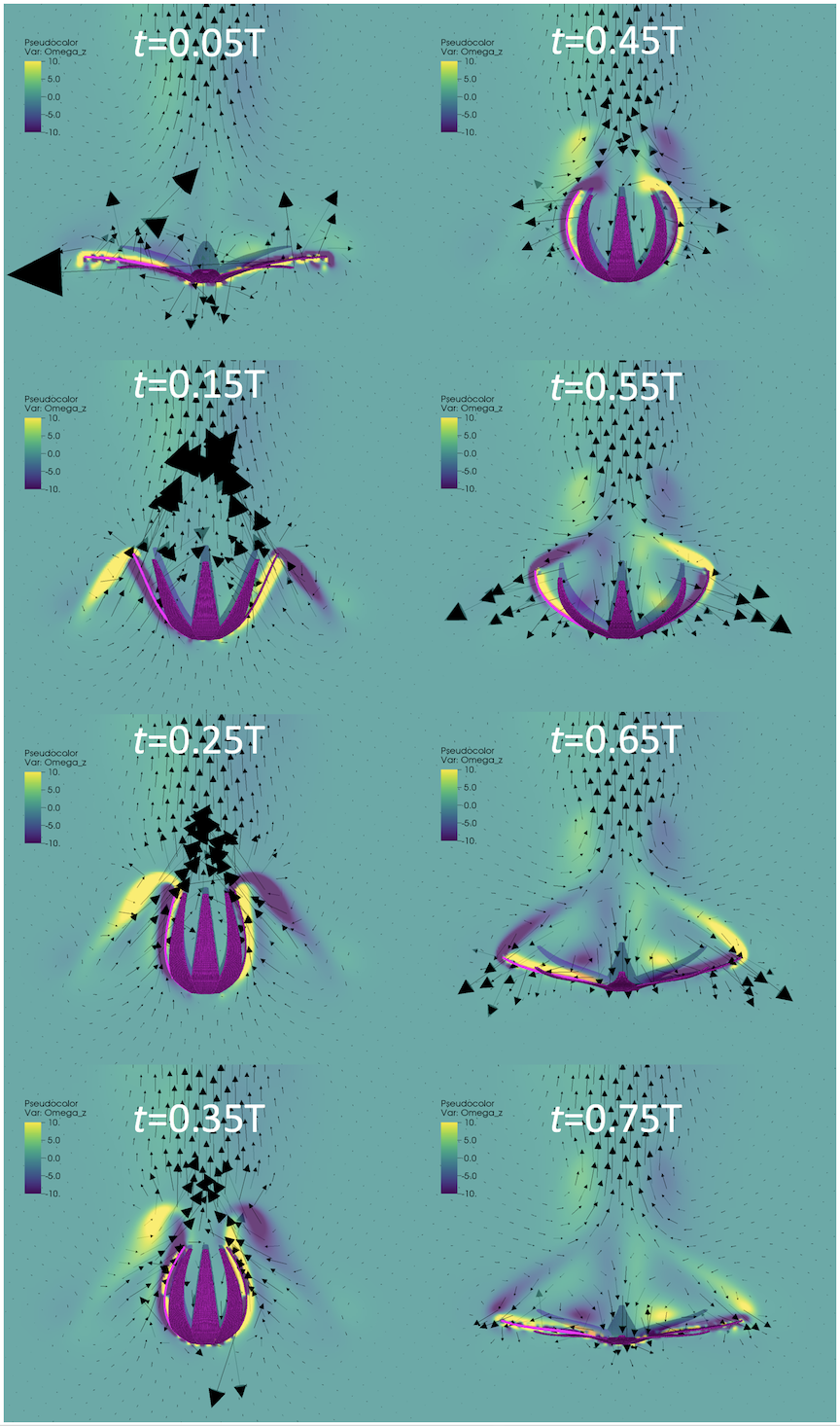}
\caption{The $z$-component of vorticity and the velocity vector field taken on a 2D plane through the central axis of the coral at $Re=80$. This $Re$ corresponds to a very large, fast pulsing coral polyp. The colormap shows the value of $\omega_z$, the arrows point in the direction of flow, and the length of the vectors correspond to the magnitude of the flow. Snapshots are taken during the fourth pulse at times that are 5\%, 15\%, 25\%, 35\%, 45\%, 55\%, 65\%, and 75\% through the cycle.}
\label{fig:Re_80_snapshots}
\end{figure}

Figures \ref{fig:Re_0p5_snapshots}-\ref{fig:Re_80_snapshots} show snapshots of the velocity and vorticity generated during the fourth pulsation cycle for three different numerical simulations corresponding to $Re=$ $0.5$, $10$, and $80$. The velocity vectors point in the direction of flow, the length of the vectors correspond to the magnitude of the flow, and the colormap corresponds to the value of  the vorticity taken in the $z$-direction (out of plane). Both vorticity and fluid velocity were taken on a 2D plane passing through the central axis of the coral polyp. The tentacles are shown in pink in 3D. The snapshots taken correspond to 5\%, 15\%, 25\%, 35\%, 45\%, 55\%, 65\%, and 75\% of the pulse such that the first three frames show the contraction phase, the next four frames show the expansion phase, and the last frame shows the polyp at rest. 

During contraction, regardless of $Re$, there is a clear upwards jet produced. In addition, vorticity is generated at the tips of the tentacles. At the beginning of expansion ($t = 0.35T$), oppositely spinning vortices are formed at the tips of each tentacle. At higher $Re$, particularly $Re=80$, the vortices formed during contraction separate from the tentacle tips and are advected upwards. The motion of these vortices help to maintain a strong upward jet above the polyp. At the lower $Re$, (e.g. $Re=0.5$), these vortices quickly dissipate. The direction of flow above the coral also reversed such that fluid is pulled downward between the tentacles. At intermediate $Re$ (e.g. $Re=10$), a weak upward jet is observed above the polyp during expansion, and fluid below this jet mixes between the tentacles.


During the resting phase (last frame), the fluid comes to rest in the lower $Re$ cases. Although during the resting phase the strength of the upwards jet in the $Re=80$ case is greatest, the magnitude of the flow between the tentacles produced by vortices generated during expansion are greater for $Re=10$. It is interesting that near the biologically relevant $Re$, the morphology of the tentacles allows for greater mixing near the polyp itself. 



\begin{figure}
\centering
\includegraphics[width=0.475\textwidth]{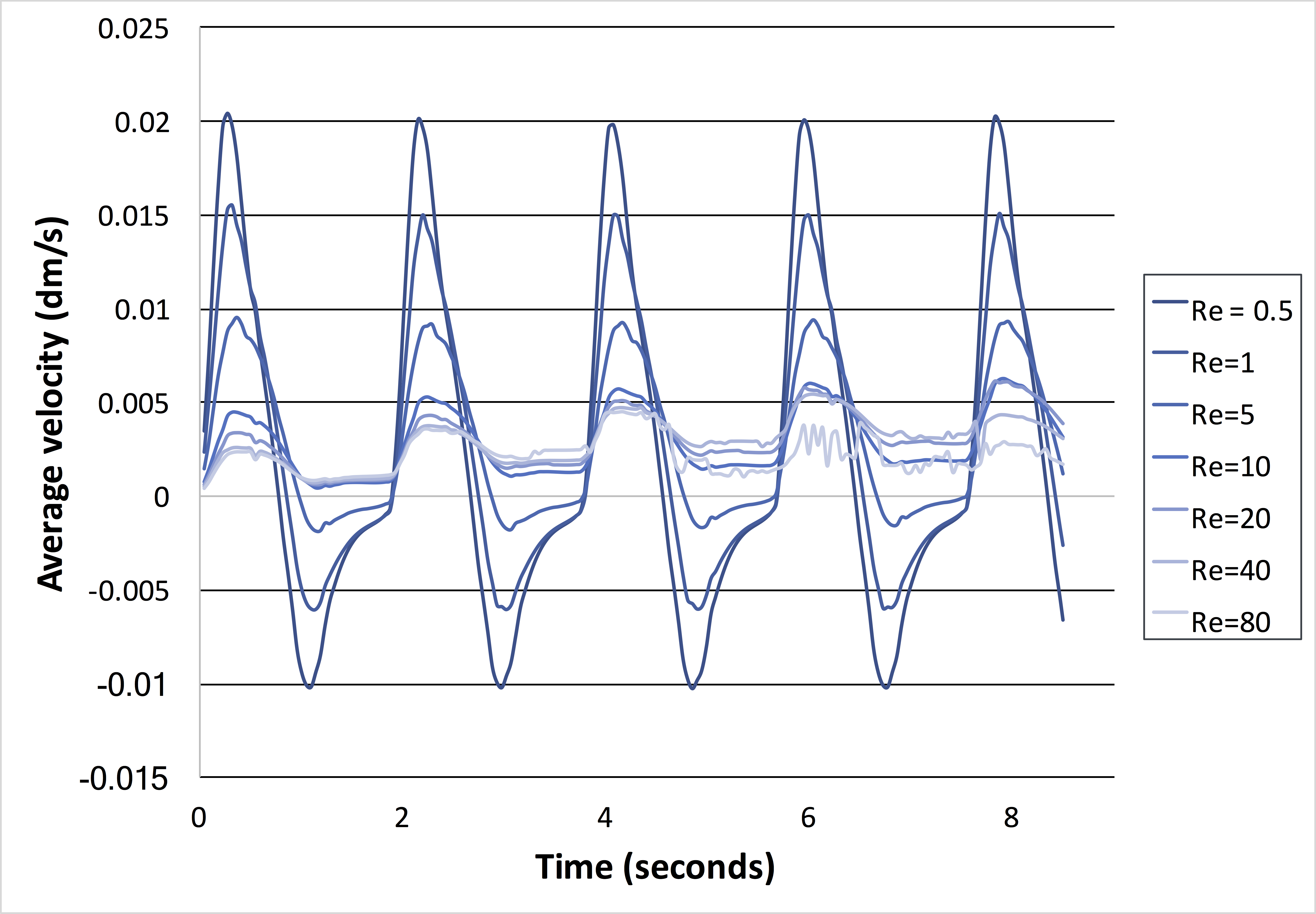}
\caption{The spatially averaged vertical flow upwards over the polyp ($u_y$) versus time for five pulse cycles. $Re=$ $0.5$, $1$, $5$, $10$, $20$, $40$, and $80$ are shown.}
\label{fig:vert_flow}
\end{figure}

\begin{figure}
\centering
\includegraphics[width=0.475\textwidth]{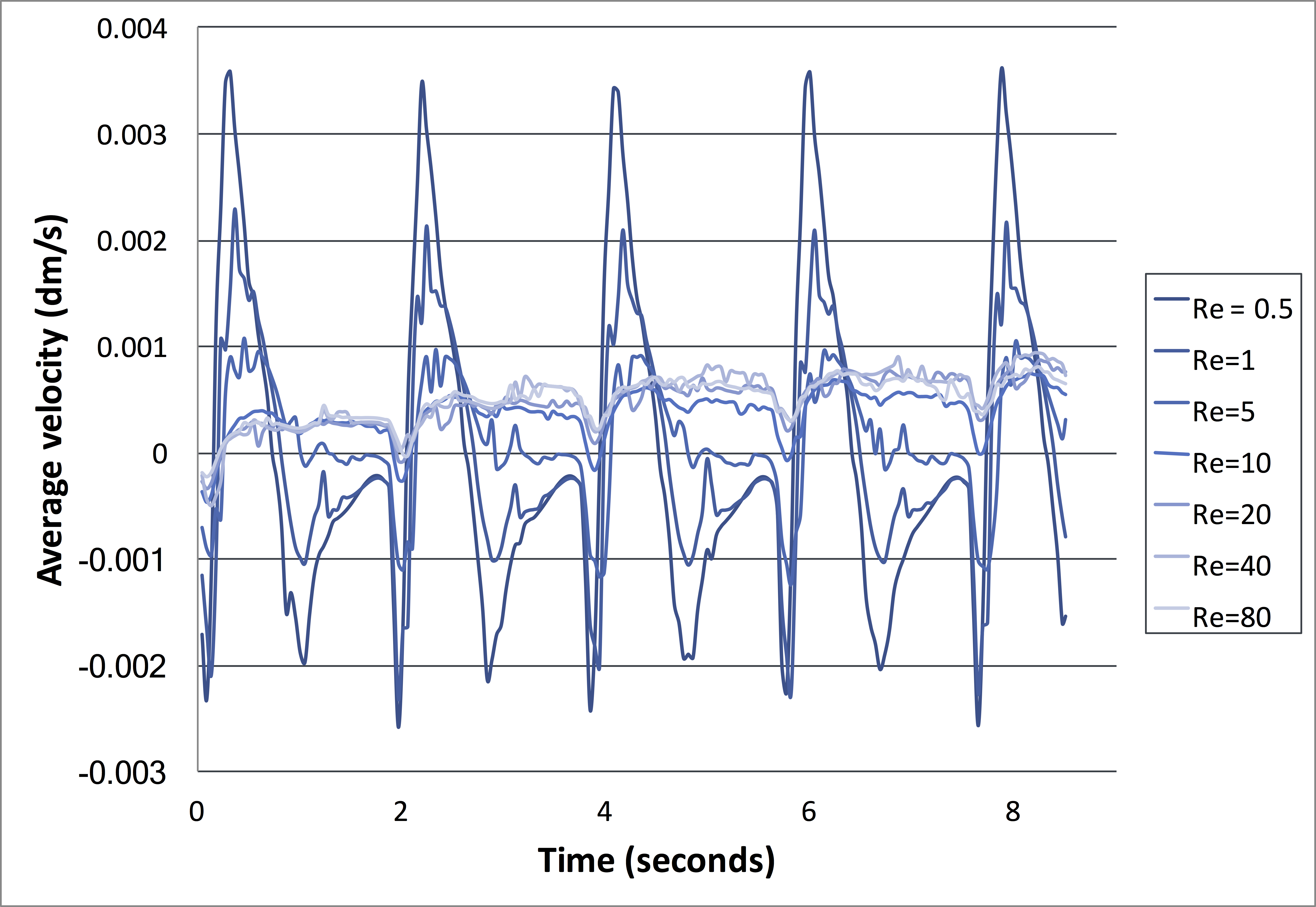}
\caption{The spatially averaged horizontal flow towards the polyp ($u_x$) over time during five pulse cycles. $Re=$ $0.5$, $1$, $5$, $10$, $20$, $40$, and $80$ are shown.}
\label{fig:hor_flow}
\end{figure}

To compare the relative strength of the upward jets generated by coral polyps across scales, we averaged the $y$-component of the velocity (in the vertical direction) within a box that was drawn from the tips of the tentacles during full contraction to one tentacle length above that point ($-0.0063 m <Y<-0.0018 m$). The width of the box was set equal to the diameter of the fully expanded polyp ($-0.0045 m<X,Z<0.0045 m$). The average vertical velocity versus time for five pulses is shown in Figure \ref{fig:vert_flow} for $Re=$ $0.5$, $1$, $5$, $10$, $20$, $40$, and $80$.

Each $Re$ investigated showed a peak average velocity in the upward jet that corresponds to the end of the contraction phase. Moreover, the largest maximal peak in average velocity corresponds to the lowest $Re=0.5$ case, while the lowest peak corresponds to the highest case, $Re=80$. This is in partially due to the fact that we average over a relatively large box, and the region of motion is larger at lower $Re$ due to the relatively large boundary layers (recall that $Re$ is lowered by increasing only dynamic viscosity). Immediately following contraction as the polyp begins to expand, the average velocity drops for each $Re$. In the cases for $Re<5$ there is significant backflow, where the average velocity becomes negative, reaches a minimum, and then slowly approaches zero. Around $Re\geq10$ there is a bifurcation, where although the average vertical flow decreases during tentacle expansion, the net average flow remains upwards. This is significant as the continuous upwards allows new fluid to be brought to the polyp throughout the pulsing cycle. 

To compare the relative strength of the flow towards the polyp, we averaged the $x$-component of the velocity (in the horizontal direction) within a box that was drawn from the tips of the tentacles during full expansion to one tentacle length to the left of that point ($-0.009 m<X<-0.0045 m$), and in the $z$-direction, the box was drawn along the diameter of the polyp fully expanded ($-0.0045 m<Z<0.0045 m$). In the vertical direction, the box was drawn from the polyp base to the top of the fully contracted tentacle ($-0.01 m<Y<-0.0063 m$). The average horizontal velocity versus time for five pulses is given in Figure \ref{fig:hor_flow} for $Re=$ $0.5$, $1$, $5$, $10$, $20$, $40$, and $80$.

For all cases of $Re$ considered as the polyp begins to contract, the average flow is away from the polyp during the first $5\%$ of the pulsation period, with the highest average velocities corresponding to the lowest $Re$, $Re=0.5$. The lowest average velocity corresponds to the highest $Re$, $Re=80$. The initial negative values are due to the whip-like motion of the tentacles at the beginning of contraction. Highest average velocities are seen at the lowest $Re$ due to the relatively larger boundary layers. After the initial contraction motion, the average velocities become positive, indicating bulk flow towards the polyp. For all $Re$, the average velocity increases until the contraction phase is over. The highest peak average velocity, again, corresponds to the lowest $Re$, $Re=0.5$; however, for $Re\geq 10$, their associated peaks of average velocity are almost equivalent. Moreover, for $Re\geq 10$, the average velocity remains towards the polyp and almost constant during the expansion and relaxation phases. At the start of the next contraction phase of the successive pulsation cycle, the average velocity dips, once again within the first $\sim5\%$ of the pulsation cycle. In contrast, for $Re\leq 5$, once the expansion phase begins, the average velocity decreases. For $Re\leq1$, the average velocity decreases, reaches a minimum, and then approaches zero. In the case of $Re=5$, during expansion the average velocity monotonically decreases toward zero before the start of the next pulsation cycle.

%
%

\section{Conclusion}
The results of this paper highlight important $Re$ bifurcations in the exchange currents generated by pulsing soft coral. At the biologically relevant $Re$, $Re=10$, the fluid-structure model suggests the polyp is able to bring nutrients or waste to absorb or exchange, respectively, towards itself. A continuous upward jet is observed throughout the entire pulsation cycle for $Re\geq 10$. This jet is significant since at these scales the polyp is able to remove waste up and away from the coral. For $Re<10$, significant backflow is observed which would result in resampling of the same fluid and reduced removal of waste. Typically pulsing corals are not observed at these scales. In terms of horizontal flow towards the polyp, this $Re$ bifurcation also has consequences for the flux of new fluid brought towards the polyp. For $Re\geq 10$, the flow continually moves towards the polyp along the substrate during most of contraction (post $5\%$ pulsation period) as well as the expansion and relaxation phases.

%
%

\section*{Acknowledgment}

The authors would like to thank Uri Shavit and Roi Holzman for introducing us to pulsing soft corals and for their assistance in the field and the organizers of the 2017 BIOMATH meeting at Kruger Park, South Africa. The authors would also like to acknowledge funding from NSF PHY grant \#1505061 (to S.K.) and \#1504777 (to L.A.M.), NSF DMS grant \#1151478 (to L.A.M.), and NSF DMS grant \#1127914 (to the Statistical and Applied Mathematical Sciences Institute). Travel support for J.E.S. was obtained from the Company of Biologists, and J.E.S. was supported by an HHMI International Student Research Fellowship and the Women Diver's Hall of Fame. 

%
%

\appendix

\section{Details on IB and IBAMR}
\label{IB_Appendix}

A three-dimensional formulation of the immersed boundary method is discussed here. For a full review of the immersed boundary method, please see Peskin \cite{Peskin:2002}. 

%
%

\subsection{Governing Equations of IB}

The governing equations for an incompressible, viscous fluid motion are given below:

\begin{align} 
   \nonumber\rho\Big[\frac{\partial\U}{\partial t}({\bf x},t) &+\U({\bf x},t)\cdot\nabla \U({\bf x},t)\Big]=  \nabla p({\bf x},t) \\ 
   &\ \ \ \ \ \ \ \ \ \ \ \ \ \ \ \ + \mu \Delta \U({\bf x},t) + \F({\bf x},t) \label{eq:NS1}
 \end{align}
  \begin{equation}
      \div \U({\bf x},t) = 0 \label{eq:NSDiv1}
  \end{equation}
where $\U({\bf x},t) $ is the fluid velocity, $p({\bf x},t) $ is the pressure, $\F({\bf x},t) $ is the force per unit area applied to the fluid by the immersed boundary, $\rho$ and $\mu$ are the fluid's density and dynamic viscosity, respectively. The independent variables are the time $t$ and the position ${\bf x}$. The variables $\U, p$, and $\F$ are all written in an Eulerian frame on the fixed Cartesian mesh, $\textbf{x}$. 

The interaction equations, which handle the communication between the Eulerian (fluid) grid and Lagrangian (boundary) grid are written as the following two integral equations:
\begin{align}
   {\bf F}({\bf x},t) &= \int {\bf f}(s,t)  \delta\left({\bf x} - {\bf X}(s,t)\right) dq \label{eq:force1} \\
   {\bf U}({\bf X}(s,t))  &= \int \U({\bf x},t)  \delta\left({\bf x} - {\bf X}(s,t)\right) d{\bf x} \label{eq:force2}
\end{align}
where ${\bf f}(s,t)$ is the force per unit length applied by the boundary to the fluid as a function of Lagrangian position, $s$, and time, $t$, $\delta({\bf x})$ is a three-dimensional delta function, and ${\bf X}(s,t)$ gives the Cartesian coordinates at time $t$ of the material point labeled by the Lagrangian parameter, $s$. The Lagrangian forcing term, ${\bf f}(s,t)$, gives the deformation forces along the boundary at the Lagrangian parameter, $s$. Equation (\ref{eq:force1}) applies this force from the immersed boundary to the fluid through the external forcing term in Equation (\ref{eq:NS1}). Equation (\ref{eq:force2}) moves the boundary at the local fluid velocity. This enforces the no-slip condition. Each integral transformation uses a three-dimensional Dirac delta function kernel, $\delta$, to convert Lagrangian variables to Eulerian variables and vice versa.

The way deformation forces are computed, e.g., the forcing term, $\textbf{f}(s,t)$, in the integrand of Equation (\ref{eq:force1}), is specific to the application. To prescribe the motion of the coral boundary, the boundary points are tethered to target points, which can be moved in a prescribed fashion. The prescribed motion of the boundary itself comes through a penalty term, tethering the Lagrangian points to the target points. The equation describing this model is
\begin{equation}
{\bf f}(s,t) = k_{targ} \left(\Y(s,t) - {\bf X}(s,t)\right),
\label{eq:force3}
\end{equation}
where $k_{targ}$ is a stiffness coefficient and $\Y(s,t)$ is the prescribed position of the target boundary. Note that $\Y(s,t)$ is a function of both the Lagrangian parameter, $s$, and time, $t$. Details on other forcing terms can be found in \cite{BattistaIB2d:2016,BattistaIB2d:2017}.

 The delta functions in these Eqs.(\ref{eq:force1}-\ref{eq:force2}) are the heart of the IB. In approximating these integral transformations, the following discretized and regularized delta functions, $\delta_h(\mathbf{x})$ \cite{Peskin:2002}, are used, 
\begin{equation}
\label{delta_h} \delta_h(\mathbf{x}) = \frac{1}{h^3} \phi\left(\frac{x}{h}\right) \phi\left(\frac{y}{h}\right) \phi\left(\frac{z}{h}\right) ,
\end{equation}
where $\phi(r)$ is defined as

\begin{equation}
\label{delta_phi} \phi(r) = \left\{ \begin{array}{c} \frac{1}{4}\left[1 + \cos\left(\frac{\pi r}{2}\right)\right] \ \ \ \ \ |r|\leq 2 \\
\mbox{ } \\
\ \ \ \ \ \ \ \ \ \ 0 \ \ \ \ \ \  \ \ \ \ \ \ \ \  \mbox{otherwise}. \end{array} \right.
\end{equation}

%
%

\subsection{Numerical Algorithm}
As stated in the main text, we impose periodic and no slip boundary conditions on the rectangular domain . To solve Equations (\ref{eq:NS1}), (\ref{eq:NSDiv1}),(\ref{eq:force1}) and (\ref{eq:force2}) we need to update the velocity, pressure, position of the boundary, and force acting on the boundary at time $n+1$ using data from time $n$. The IB does this in the following steps \cite{Peskin:2002}, with an additional step ($4b$) for IBAMR \cite{GriffithThesis:2005,BGriffithIBAMR}:

\textbf{Step 1:} Find the force density, ${\bf{F}}^{n}$ on the immersed boundary, from the current boundary configuration, ${\bf{X}}^{n}$.\\
\indent\textbf{Step 2:} Use Equation (\ref{eq:force1}) to spread this boundary force from the Lagrangian boundary mesh to the Eulerian fluid lattice points.\\
\indent\textbf{Step 3:} Solve the Navier-Stokes equations, Equations (\ref{eq:NS1}) and (\ref{eq:NSDiv1}), on the Eulerian grid. Upon doing so, we are updating ${\bf{u}}^{n+1}$ and $p^{n+1}$ from ${\bf{u}}^{n}$, $p^{n}$, and ${\bf{f}}^{n}$. Note that a staggered grid projection scheme is used to perform this update.\\
\indent\textbf{Step 4:}
(\emph{4a}) Update the material positions, ${\bf{X}}^{n+1}$,  using the local fluid velocities, ${\bf{U}}^{n+1}$, computed from  ${\bf{u}}^{n+1}$ and Equation (\ref{eq:force2}).
(\emph{4b}) If on a selected time-step for adaptive mesh refinement, refine the Eulerian grid in areas of the domain that contain the immersed structure or where the vorticity exceeds a predetermined threshold, . 

We note that Step 4b is from the IBAMR implementation of IB. IBAMR is an IB framework written in C++ that provides discretization and solver infrastructure for partial differential equations on block-structured locally refined Eulerian grids \cite{MJBerger84,MJBerger89} and on Lagrangian meshes. Adaptive mesh refinement (AMR) achieves higher accuracy between the Lagrangian and Eulerian mesh by increasing grid resolution in areas of the domain where the vorticity exceeds a certain threshold and in areas of the domain that contain an immersed boundary. AMR improves the computational efficiency by decreasing grid resolution in areas that do not necessitate high resolution.

The Eulerian grid was locally refined near both the immersed boundaries and regions of vorticity where $|\omega| > 0.50$. This Cartesian grid was structured as a hierarchy of four nested grid levels where the finest resolved grid was assigned a resolution of $dx = D/1024$, see Table \ref{table:num_param}. A 1:4 spatial step size ratio was used between each successive grid refinements. The Lagrangian spatial step resolution was chosen to be twice the resolution of the finest Eulerian grid, with $ds =D/2048$.

%
%

\bibliographystyle{elsarticle-num}
\bibliography{coral}

\end{document}